\documentclass[11pt]{article}
\usepackage{amssymb,color}
\newcommand{\N}{N\raise.7ex\hbox{\underline{$\circ $}}$\;$}

\begin{document}
\title{
    Confluent Heun functions and the Coulomb problem \\
    for  spin 1/2 particle  in Minkowski space}

\author{V. Balan\footnote{University Politehnica of Bucharest, Romania, vladimir.balan@upb.ro},
 A.M. Manukyan\footnote{Institute for Physical Research of NAS of Armenia},
E.M. Ovsiyuk\footnote{Mosyr State Pedagogical University, Belarus, e.ovsiyuk@mail.ru},
 V.M. Red'kov\footnote{B.I. Stepanov Institute of Physics, NAS of Belarus,
redkov@dragon.bas-net.by},
O.V. Veko\footnote{Kalinkovichi Gymnasium, Belarus,vekoolga@mail.ru}}

\maketitle

\begin{abstract}
In the paper, the well-known quantum mechanical problem of a spin 1/2 particle in external Coulomb
 potential, reduced to a system of two first-order differential equations, is studied from the point
 of view of possible applications of the Heun function theory to treat this system.
It is shown that in addition to the standard way to solve the problem in terms of the
 confluent hypergeometric functions (proposed in 1928 by G. Darvin and W. Gordon), there are
 possible several other possibilities which rely on applying the confluent Heun functions.
Namely, in the paper there are elaborated two combined possibilities to construct solutions:
 the first applies when one equation of the pair of relevant functions is expressed trough
 hypergeometric functions, and another constructed in terms of confluent Heun functions.
In this respect, certain relations between the two classes of functions are established.
It is shown that both functions of the system may be expressed in terms of confluent Heun functions.
All the ways to study this problem lead us to a single energy spectrum, which indicates their correctness.

\end{abstract}
{\small
}
\section{Introduction}
The general Heun equation is a second order linear differential equation which has four
 regular singularities and different confluent forms \cite{Heun-1889, Ronveaux-1995}.
The general Heun equation and all its confluent forms turn out to be of
 primary significance in physical applications, for instance in quantum mechanics
 and field theory on the background of curved space-time models, and in optics -- see
 [3--72].
A more complete and comprehensive list of references can be found on the site
 of {\em the Heun Project}\footnote{Sofia University, The Heun Project:
 Heun functions, their generalizations and applications, http://theheunproject.org}.\par
In this paper, the well-known quantum mechanical problem of a spin 1/2 particle in external
 Coulomb potential, reduced to a system of two first-order differential equations,
 is studied from the point of view of possible applications of the Heun function theory
 to treat this system.
It is shown that in addition to the standard way \cite{Landau-4} to solve the problem in
 terms of the confluent hypergeometric functions (proposed in 1928 by G. Darvin and W. Gordon),
 there are possible several other possibilities which rely on applying confluent Heun functions.
Namely, we shall elaborate two combined possibilities to construct solutions: one for the case
 when one equation of the pair of relevant functions is expressed trough hypergeometric functions,
 and another constructed in terms of confluent Heun functions.
As well, certain relations between these two classes of functions are established.
There exists a possibility to express both the functions of the system in terms of confluent
 Heun functions.
All the ways to study the problem lead us to a unique energy spectrum, which indicates
 their correctness. In particular physical problems, there exist as well similar
 possibilities to avoid (with the help of special tricks) the use of Heun functions by
 restricting to the use of only hypergeometric functions  -- for instance see in
 \cite{Bogush-Krylov-Ovsiyuk-Red'kov-2010,Red'kov-Ovsiyuk-2011, Red'kov-Ovsiyuk-Veko-2012,
 Ovsiyuk-Veko-Red'kov-2013}.
\section{The Coulomb problem: solutions constructed by hypergeometric and partially by
    Heun functions}
In spherical coordinates, a diagonal tetrad has the form \cite{Book-2012, Book-2014}
\begin{eqnarray} dS^{2} = dt^{2} - d r^{2} - r^{2} ( d \theta^{2} +
\sin^{2} d \phi^{2}) \;, \nonumber \\
e_{(0)}^{\alpha} = ( 1,0,0,0) \;, \qquad e_{(3)}^{\alpha} = ( 0,
1,0,0) \;, \;
\nonumber
\\
e_{(1)}^{\alpha} = ( 0,0, r^{-1}, 0) \;, \qquad e_{(2)}^{\alpha}
= ( 0,0, 0, r^{-1} \sin^{-1} \theta ) \;. \label{1.6.1}
\end{eqnarray}
The Ricci rotation coefficients are
$$
\gamma_{ab0} = 0, \; \gamma_{ab3} = 0, \quad \gamma_{ab1}
= \left | \begin{array}{cccc}
0 & 0 & 0 & 0 \\
0 & 0 & 0 & - r^{-1} \\
0 & 0 & 0 & 0 \\
0 & + r^{-1} & 0 & 0
\end{array} \right |, \;
\gamma_{ab2} = \left | \begin{array}{cccc}
0 & 0 & 0 & 0 \\
0 & 0 & \cot \theta r^{-1}& 0 \\
0 & - \cot \theta r^{-1} & 0 & - r^{-1} \\
0 & 0 & + r^{-1} \chi & 0
\end{array} \right |.
$$
Generally, the covariant Dirac equation\footnote{We further use the notations from
    \cite{Book-2012, Book-2014}.}
$$
\left [ i \gamma^{c} \Big( e_{(c)}^{\alpha} \partial _{\alpha} +
{1 \over 2} j^{ab} \gamma_{abc}\Big) -m \right ] \Psi = 0
$$
takes the form
\begin{eqnarray}
\left [ i \gamma^{0} {\partial \over \partial t } + i
\Big(\gamma^{3} {\partial \over \partial r} + {\gamma^{1}
j^{31} + \gamma^{2} j^{32} \over r }\Big) +
{1 \over r } \Sigma_{\theta \phi} - m \right ] \Psi
= 0, \label{1.6.2a}
\end{eqnarray}
\begin{eqnarray}
\Sigma _{\theta,\phi } = i \gamma ^{1} \partial
_{\theta} + \gamma ^{2} {i \partial _{\phi} \; +
 i \sigma ^{12} \over \sin \theta }.
\label{1.6.2b}
\end{eqnarray}
With the help of relations
$$
{\gamma^{1} j^{31} + \gamma^{2} j^{32} \over r } =
{ \gamma^{3} \over r }, \quad \Psi = {1
\over r } \tilde{\Psi}
$$
equation (\ref{1.6.2a}) is simplified,
\begin{eqnarray}
\left ( i \gamma^{0} {\partial \over \partial t } + i
\gamma^{3} {\partial \over \partial r} + {1 \over r }
\Sigma_{\theta \phi} - m \right ) \tilde{\Psi} = 0
. \label{1.6.2c}
\end{eqnarray}
In order to diagonalize the operators $i\partial_{t}, \vec{J}^{2},J_{3}$, one takes
 the wave function in the form \cite{Book-2012}
\begin{eqnarray}
\tilde{\Psi} = e^{-i E t } \left | \begin{array}{r}
f_{1}(\chi) \; D_{-1/2} \\
f_{2}(\chi) \; D_{+1/2} \\
f_{3}(\chi) \; D_{-1/2} \\
f_{4}(\chi) \; D_{+1/2}
\end{array} \right |,
\label{1.6.3}
\end{eqnarray}
where the Wigner functions \cite{Varshalovich-Moskalev-Hersonskiy-1975} are denoted by
    $D_{\sigma}=D^{j}_{-m.\sigma}(\phi, \theta.0)$.
After separating the variables, we get four radial equations (let $ \nu =j+1/2$)
\begin{eqnarray}
E f_{3} - i {d \over d \chi} f_{3} - i {\nu \over
\sin \chi } f_{4} - m f_{1} = 0, \quad E
f_{4} + i {d \over d \chi } f_{4} + i {\nu \over \sin \chi}
f_{3} - m f_{2} = 0 ,
\nonumber
\\
E f_{1} + i {d \over d \chi } f_{1} + i {\nu \over
\sin \chi } f_{2} - m f_{3} = 0, \quad
 E
f_{2} - i {d \over d \chi } f_{2} - i {\nu \over \sin \chi }
f_{1} - m f_{4} = 0.
\label{1.6.4}
\end{eqnarray}
In a spherical tetrad, the space reflection operator is given by \cite{Book-2012}
$$
\hat{\Pi}_{sph} \; \; = \left | \begin{array}{cccc}
0 & 0 & 0 & -1 \\
0 & 0 & -1 & 0 \\
0 & -1& 0 & 0 \\
-1& 0 & 0 & 0
\end{array} \right |
\; \otimes \; \hat{P} \;.
$$
From the eigenvalues equations $\hat{\Pi}_{sph}\; \Psi_{jm} = \; \Pi \; \Psi _{jm}$, we obtain
\begin{eqnarray}
\Pi = \; \delta \; (-1)^{j+1}, \;\; \delta = \pm 1 \;, \qquad
f_{4} = \; \delta \; f_{1}, \qquad f_{3} = \;\delta \; f_{2} \;,
\label{1.6.5}
\end{eqnarray}
which simplifies (\ref{1.6.4}),
\begin{eqnarray}
\Big({d \over d r } + {\nu \over r } \Big) \; f +( E + \delta m )\; g = 0, \quad
\Big({d \over d r } - {\nu \over r } \Big)\; g -( E - \delta m ) \; f = 0 ,
\label{1.6.6}
\end{eqnarray}
where instead of $f_{1}$ and $f_{2}$, the new variables $f$ and $g$ are used
$$
f = {f_{1} + f_{2} \over \sqrt{2}}, \qquad g = {f_{1} -
f_{2} \over i \sqrt{2}}.
$$
For definiteness, let us consider the case when $\delta =1$,
\begin{eqnarray}
\Big({d \over d r } + {\nu \over r } \Big) f +( E + m ) g = 0, \quad
\Big({d \over d r } - {\nu \over r } \Big) g -( E - m ) f = 0  ;
\label{1.6.6'}
\end{eqnarray}
by performing the replacement $m\rightsquigarrow -m $ we obtain the equations for $\delta =-1$.\par
The presence of the external Coulomb field is taken into account in (\ref{1.6.6'}) by the formal change
    $\epsilon \rightsquigarrow \epsilon + {e \over r}.$
Thus, the quantum Coulomb problem for a Dirac particle is described by the following radial system
\begin{eqnarray}
 \Big( {d \over d r } + {\nu \over r } \Big) f + \Big( E + { e \over r } + m \Big) g = 0,\quad
 \Big({d \over d r } - {\nu \over r } \Big) g - \Big( E + {e \over r } - m \Big) f = 0 .
\label{1.10.1}
\end{eqnarray}
Let us perform a linear transformation over the functions $f(r)$ and
 $g(r)$\footnote{The coefficients of this transformation may depend
 on the radial variable; let its determinant obey the identity
 $a (r) b (r) - c (r) d(r) = 1$.}
\begin{eqnarray}
f (r) = a F (r) + c G (r), \quad g (r) = d F(r) + b G(r) \;,\nonumber\\
F(r) = b f(r) - c g (r), \quad G(r) = - d f(r) + a g(r).
\label{1.10.2}
\end{eqnarray}
Let us combine the equations (\ref{1.10.1}) as follows: the first equation is
 multiplied by $+b$, the second by $- c$, and then sum the results;
 analogously, we add the first equation multiplied by $-d$ with the second
 multiplied by $+a$. Thus we arrive at
\begin{eqnarray}
\left [ \; {d \over d r} -b'a +c'd + {\nu \over r} \; (ba +cd) +\Big(E+{e \over r}
+ m\Big) \; bd + \Big(E + {e \over r} -m\Big)\;ca \; \right ] F\nonumber\\
= \left [ \; b'c - b c' - {\nu \over r} 2bc - \Big(E+{e \over r}
+m\Big) b^{2} - \Big(E + {e \over r} -m\Big) c^{2} \; \right ] G\;,\nonumber\\
\left [ \; {d \over d r} + d'c - a'b - {\nu \over r} \; (dc
+ab) - \Big(E+{e \over r} + m\Big) \; bd - \Big(E + {e \over r}-m\Big)
\; ca \; \right ] G\nonumber\\
= \left [ \; -d'a +d a' + {\nu \over r} 2ad + \Big(E+{e \over r}
+m\Big) d^{2} + \Big(E + {e \over r} -m\Big) a^{2} \; \right ] F\;.
\label{1.10.3}
\end{eqnarray}
For simplicity, let us assume that the transformation (\ref{1.10.2}) does not depend on
 $r$, and let it be orthogonal:
\begin{eqnarray}
S = \left | \begin{array}{cc}a & c \\d & b\end{array} \right | =
\left | \begin{array}{rr}\cos A/2 & \sin A/2 \\- \sin A/2 & \cos A/2\end{array} \right |,
\label{1.10.4}
\end{eqnarray}
which simplifies the equations (\ref{1.10.3})
\begin{eqnarray}
\left ( {d \over d r} + {\nu \over r} \cos A - m \sin A\right ) F = \left (
- {\nu \over r} \sin A -{e \over r} - E - m \cos A\right ) G,\nonumber\\
\left ( {d \over d r}  - {\nu \over r} \cos A + m\sin A \right ) G =
\left ( - {\nu \over r} \sin A +{e \over r} + E - m \cos A \right ) F.
\label{1.10.5}
\end{eqnarray}
There exist four possibilities (only two of them are different in fact):\par\vspace{2mm}
1)
$$
- {\nu \over r} \sin A +{e \over r}=0\;, \qquad \sin A = {e \over \nu }\;,
\qquad\cos A = \sqrt{1 - e^{2} / \nu^{2}}\;,
$$
$$
\cos{A\over 2} = \sqrt{{\nu + \sqrt{\nu^{2} - e^{2}} \over 2\nu}}\;, \qquad
\sin{A\over 2} = \sqrt{{\nu - \sqrt{\nu^{2} -e^{2}} \over 2 \nu}} \; ;
$$
\indent
$1'$)
\begin{eqnarray}
 - {\nu \over r} \sin A -{e \over r}=0\;, \qquad \sin A = - {e \over \nu },
 \quad \cos A = \sqrt{1 - e^{2} / \nu^{2}},\nonumber\\
\cos{A\over 2} = \sqrt{{\nu -\sqrt{\nu^{2} - e^{2}} \over 2
\nu}}, \quad \sin{A\over 2} = \sqrt{{\nu + \sqrt{\nu^{2} -e^{2}} \over 2 \nu}} \; ;
\label{1.10.6a}
\end{eqnarray}
\indent
2)
$$
E - m \cos A = 0, \quad \cos A = + {E \over m}, \quad \sin A = \sqrt{1 -E^{2} / m^{2} },
$$
$$
\cos {A\over 2} = \sqrt{{m+E \over 2m}}, \qquad \sin {A\over 2} = \sqrt{{m-E \over 2m}} \; ;
$$
\indent
$2'$)
\begin{eqnarray}
- E - m \cos A = 0, \qquad \cos A = -{E \over m}, \quad
\sin A = \sqrt{1 -E^{2} / m^{2} },\nonumber\\
\cos {A\over 2} = \sqrt{{m-E \over 2m}}, \quad \sin {A\over 2}
= \sqrt{{m+E \over 2m}} \;. \label{1.10.6b}
\end{eqnarray}
\indent
First, consider the case 1). Equations (\ref{1.10.5}) take the form
\begin{eqnarray}
\left ( {d \over d r} + {\nu \over r} \cos A - m \sin A\right )
 F = \left ( -{2e \over r} - E - m \cos A \right ) G,\nonumber\\
\left ( {d \over d r}  - {\nu \over r} \cos A + m \sin A
\right ) G = ( E - m \cos A ) F.
\label{1.10.7}
\end{eqnarray}
After eliminating the function $F$, we get a second order equation for $G$
\begin{eqnarray}
\left ( {d \over d r} + {\nu \over r} \cos A - m \sin A
\right ) \left ( {d \over d r}  - {\nu \over r} \cos A + m
\sin A \right ) G\nonumber\\
= ( E - m \cos A ) \left ( -{2e \over r} - E - m \cos A\right ) G,
\label{1.10.8a}
\end{eqnarray}
or, differently,
\begin{eqnarray}
\left ( {d^{2} \over dr^{2}} + E^{2} -m^{2} + {\nu \cos A -
\nu^{2} \cos^{2} A \over r^{2} } + {2eE -
 2e m\cos A +2m\nu \sin A \cos A \over r}  \right ) G = 0.\nonumber
\end{eqnarray}
Having in mind the identity $\sin A = e/ \nu$, the last equation reduces to
\begin{eqnarray}
\left (\; {d^{2} \over dr^{2}} + E^{2} -m^{2} + {\nu \cos A -
\nu^{2} \cos^{2} A \over r^{2} } + {2eE \over r}  \right ) G =0.
\label{1.10.8b}
\end{eqnarray}
After changing the variable, $x=2\,\sqrt{m^{2}-E^{2}}\,r$, it reads
\begin{eqnarray}
{d^{2}G\over dx^{2}} + \left(-{1\over 4}-{\nu \cos A\, (\nu \cos
A-1)\over x^{2}} + {e E \over \sqrt{m^{2}-E^{2}}\,x}\right)G=0\,.
\label{1.10.8b'}
\end{eqnarray}
With the use of the substitution $ G (x) = x^{a} e^{bx}\bar{G} (x)$ for $\bar{G}$, we get
$$
x\,{d^{2}\bar{G}\over dx^{2}}+(2\,a+2\,b\,x)\,{d \bar{G}\over dx}
$$
$$
+\left[(b^{2}-{1\over 4})\,x+{a^{2}-a-\nu \cos A\ (\nu \cos
A-1)\over x}+2ab+ {eE \over \sqrt{m^{2}-E^2}}\right]\bar{G}=0\,.
$$
When
$$
a = +\nu\,\cos A = \sqrt{\nu^{2} - e^{2}} \,, \qquad b=-{1 \over 2}\;,
$$
this equation for $\bar{G}$ becomes simpler,
$$
x\,{d^{2}\bar{G}\over dx^{2}}+(2\,a-\,x)\,{d \bar{G}\over dx}
-\left(a-{eE \over \sqrt{m^{2}-E^2}}\right)\varphi=0\,,
$$
which is a confluent hypergeometric equation
$$
x \; _{1} F _{1}''(x) +( \gamma -x) _{1} F _{1}' (x) - \alpha \; _{1} F _{1}(x) =0 \;, \qquad \alpha=
a-{eE \over \sqrt{m^{2}-E^2}}\,, \qquad \gamma=2a\,.
$$
To polynomials there corresponds the known restriction $\alpha = - n$, $n=0,1,2,...$,
 which gives the known energy quantization rule
\begin{eqnarray}
a-{e E \over \sqrt{m^{2}-E^2}} = -n \qquad \Longrightarrow \qquad E
= { m \over \sqrt{1 + e^{2} / (n + \sqrt{\nu^{2} - e^{2}} )^{2}}} \;.
\label{1.10.8c}
\end{eqnarray}
In turn, from (\ref{1.10.7}) it follows a second order equation for $F$,
$$
\left ( {d \over d r}  - {\nu \over r} \; \cos A + m \;\sin
A \right ) {r \over 2e + (E + m \cos A )r } \left ( {d
\over d r} + {\nu \over r} \;\cos A - m \; \sin A \right )F
$$
\begin{eqnarray}
 = ( -E + m \cos A ) F \;,
\label{1.10.9a}
\end{eqnarray}
or
$$
 \left [ \Big( {d \over d r}  - {\nu \over r} \; \cos A + m \;\sin A \Big)
\Big( {d \over d r} + {\nu \over r} \;\cos A - m \; \sin A\Big) \right.
$$
$$
\left. + {d \over dr } \ln \left ( {r \over 2e + (E + m
\cos A )r } \right ) \Big( {d \over d r} + {\nu \over r} \;\cos
A - m \; \sin A \Big)\right ] F
$$
$$
=\Big( { 2e \over r} + E + m \cos A \Big)\; \Big( -E + m \cos A \Big) \; F \;,
$$
and further,
$$
\left [ {d^{2} \over dr^{2}} + \left ({1 \over r} - {E+ m \cos A
\over 2e +(E+ m \cos A) r} \right ) {d \over d r} \right.
$$
$$
\left. + {\nu \over r^{2} } \;\cos A - {m \; \sin A \over r} -
\left ( {1 \over r} -{2e \over r [ 2e +(E+ m \cos A) r]} \right )
\Big( {\nu \over r} \;\cos A - m \; \sin A \Big) \right.
$$
$$
\left. + E^{2} -m^{2} - {\nu \cos A \over r^{2}} + {e^{2} - \nu^{2}
\over r^{2}}  +{2eE \over r}  \right ] F = 0 \;.
$$
Finally, we obtain
$$
\left [ {d^{2} \over dr^{2}} + \Big({1 \over r} - {E+ m \cos A
\over 2e +(E+ m \cos A) r} \Big) {d \over d r} \right.
$$
$$
\left. + E^{2} -m^{2} +{2eE \over r} + {e^{2} - \nu^{2}\over r^{2}} -
 { 2em \sin A + \nu \cos A (E +m \cos A) \over r \; [ 2e + (E+ m \cos A) r] }
 \right ] F = 0 \;.
$$
\indent
Let us introduce special designation for the additional singular point
$$
-{2e \over E + m \cos A } = R \; ;
$$
then we obtain
\begin{eqnarray}
\left [ {d^{2} \over dr^{2}} + \left({1 \over r} - {1 \over r - R} \right) {d \over d r}
 + E^{2} -m^{2} +{ 2eE \over r} + {e^{2} - \nu^{2}\over r^{2}} +
 { m R \; \sin A - \nu \cos A \over r \; ( r-R) }\right ] F = 0.
\label{1.10.9b}
\end{eqnarray}
After changing the variable $ y = r / R$, it reads
\begin{eqnarray}
{d^{2}F\over dy^{2}}+\left ({1\over y}-{1\over
y-1}\right ) \,{dF\over dy} +\left ( (E^{2}-m^{2})\,R^{2}-
{\nu^{2}-e^{2}\over y^{2}} \right.\nonumber\\
\left. + {-\nu \cos A+m R\, \sin A\over y-1}+ { 2eRE-m R\sin
A+\nu \cos A \over y} \right )F=0\,. \label{1.10.9c}
\end{eqnarray}
\indent
Let us search solutions in the form
 $F=y^{a}\,e^{by}\,\bar{F}(y)$; the function $\bar{F}$ obeys
$$
{d^{2} \bar{F} \over dy^{2}}+\left({2a+1\over y}+2b-{1\over
y-1}\right)\,{d \bar{F}\over dy}
$$
$$
+ \left[b^{2}+(E^{2}-m^{2})\,R^{2}+ {a^{2}-\nu^{2}+e^{2}\over
y^{2}}-{a+b+\nu \cos A-m R\, \sin A\over y-1} \right.
$$
$$
\left. + {a+b+2ab+R(2eE-m \sin A)+\nu \cos A\over y}\right]\,\bar{F}=0\,.
$$
With $a,b$ taken according to\footnote{We shall use below underlined values.}:
\begin{eqnarray}
a=\underline{+\sqrt{\nu^{2}-e^{2}}}\,, \;\; -\sqrt{\nu^{2}-e^{2}}\;,\nonumber\\
b= +\sqrt{m^{2}-E^{2}}\,R\,, \underline{ -\sqrt{m^{2}-E^{2}}\,R}
\label{1.10.10a}
\end{eqnarray}
the above equation becomes simpler
\begin{eqnarray}
{d^{2} \bar{F} \over dy^{2}}+\left (2b+{2a+1\over y}-{1\over
y-1}\right )\,{d \bar{F}\over dy}\nonumber\\
+\left ({a+b+2ab+ 2eRE- m R\sin A + \nu \cos A\over y}\right.\nonumber\\
\left.-{a+b+\nu\cos A-m R\, \sin A\over y-1}\right )\,\bar{F}=0\,.
\label{1.10.10b}
\end{eqnarray}
This can be easily recognized as a confluent Heun equation for
 $H(\alpha,\beta,\gamma,\delta,\eta,z)$,
\begin{eqnarray}
H''+\left ( \alpha+{1+\beta\over z}+{1+\gamma\over z-1}\right )H'
+ \left ({1\over 2}\,{\alpha+\alpha \beta-\beta-\beta
\gamma-\gamma-2\eta\over z}\right.\nonumber\\
\left.+{1\over 2}\, {\alpha+\alpha
\gamma+\beta+\beta \gamma+\gamma+ 2\delta+2\eta\over z-1}\right )H=0\,
\label{1.10.10c}
\end{eqnarray}
with the parameters
\begin{eqnarray}
\alpha=2b\,, \qquad \beta=2a\,, \qquad \gamma=-2\,,\nonumber\\
\delta=2\,e ER\,, \qquad \eta=1+m R\,\sin A-2\,e ER-\nu \cos A\,.
\label{1.10.10d}
\end{eqnarray}
Let us use the known condition  \cite{Ronveaux-1995}
    (one of two conditions needed to get polynomials):
\begin{eqnarray}
\delta=-\left (n+{\beta+\gamma+2\over 2}\right )\,\alpha\,,\qquad n=0,1,2,\dots,
\label{1.10.11a}
\end{eqnarray}
it results the energy quantization rule
$$
a= + \sqrt{\nu^{2}-e^{2}}\,,\qquad b= - \sqrt{m^{2}-E^{2}}\,R\,,
$$
$$
eER=(n+\sqrt{\nu^{2}-e^{2}})\,\sqrt{m^{2}-E^{2}} \; R \;,
$$
so we infer
\begin{eqnarray}
E = { m \over \sqrt{1 + e^{2} / (n + \sqrt{\nu^{2} - e^{2}})^{2}  }} \;, \label{1.10.11b}
\end{eqnarray}
which coincides with the known formula for energy levels.\par
It should be emphasized that as follows from (\ref{1.10.7}), the function
 $F$ (being constructed in terms of the confluent Heun functions) can be
 related with the function $G$ (which is determined in terms of confluent
 hypergeometric functions) by means of the following differential operators:
\begin{eqnarray}
G= \left ( -{2e \over r} - E - m \cos A \right )^{-1} \left ( {d \over d r}
 + {\nu \over r} \cos A - m \sin A\right ) F,
\label{1.10.7'}
\end{eqnarray}
and
\begin{eqnarray}
F = {1 \over ( E - m \cos A ) } \left ( {d \over d r}  - {\nu \over r} \cos A
+ m \sin A\right ) G.
\label{1.10.7'x}
\end{eqnarray}
They can be rewritten in the form
\begin{eqnarray}
G (r) = {1 \over 2e}{ y \over y -1 } \left ( {d \over d y} + {\nu \over y}
\cos A - m R  \sin A\right ) F (y),
\label{1.10.7a}
\end{eqnarray}
and
\begin{eqnarray}
F (y) = {r \over 2( E r + e )} \left ( {d \over d r}  - {\nu \over r} \cos A + m \sin A\right ) G (r).
\label{1.10.7b}
\end{eqnarray}
\vspace{5mm}\indent
Let us consider the case 2) -- see (\ref{1.10.6b}).
The equations (\ref{1.10.5}) take the form
\begin{eqnarray}
\left ( {d \over d r} + {\nu \over r} \;\cos A - m \; \sin A\right )
 F = \left ( - {\nu \sin A + e \over r}  - 2m \cos A \right ) G \;,\nonumber\\
\left ( {d \over d r}  - {\nu \over r} \; \cos A + m \;
\sin A \right ) G =  {e-\nu \sin A \over r} \; F \;.
\label{1.10.12}
\end{eqnarray}
One can obtain a second order equation for $G(r)$:
\begin{eqnarray}
\left [ \Big( {d \over d r} + {\nu \over r} \;\cos A - m \;
\sin A \Big) \;r \Big( {d \over d r}  - {\nu \over r} \;
\cos A + m \; \sin A \Big) \right.\nonumber\\
\left. + ( e - \nu \sin A ) \Big( { e + \nu \sin A \over r} +
2m \cos A \Big) \right ] G = 0 \;,
\label{1.10.13a}
\end{eqnarray}
from where it follows
$$
\left [ {d \over d r}  - {\nu \over r} \; \cos A + m \;\sin
A + r \left ( {d ^{2}\over d r^{2} } + {\nu \cos A \over r^{2} } \; -
 \Big( {\nu \over r} \;\cos A - m \; \sin A \Big)^{2} \right ) \right.
$$
$$
\left. + ( e - \nu \sin A ) \left ( { e + \nu \sin A \over r }
+ 2m \cos A \right ) \right ] G = 0 \;,
$$
or
$$
\left ( {d ^{2}\over d r^{2} } + {1 \over r } {d \over d r}
- m ^{2} \; \sin^{2} A   +  { e^{2} - \nu ^{2} \over r^{2} } +
 { 2me \cos A \over r} + {m \sin A \over r } \right ) G = 0 \;.
$$
Further, taking into account the identity $\cos A = E /m$, we get
\begin{eqnarray}
\left ( {d ^{2}\over d r^{2} }   + {1 \over r } {d \over d r}  +
 E^{2} - m^{2}   +  { e^{2} - \nu ^{2} \over r^{2} } +
 { 2 e E \over r} + {\sqrt{m^{2} - E^{2} } \over r } \right ) G = 0 \;.
\label{1.10.13b}
\end{eqnarray}
By making the change of variables $x=2\,\sqrt{m^{2}-E^{2}}\,r$, we get
$$
{d^{2}G\over dx^{2}} +{1\over x}\,{dG\over dx}+ \left(-{1\over
4}-{\nu^{2}-e^{2}\over x^{2}} +{1\over 2}\, {m^{2}-E^{2}+2 E
e\,\sqrt{m^{2}-E^{2}} \over (m^{2}-E^{2})\,x}\right)G=0\,.
$$
Let $ G (x) = x^{a} e^{bx} \bar{G}(x)$; the function $\bar{G}$ satisfies
$$
x\,{d^{2}\bar{G}\over dx^{2}}+(2\,a+1+2\,b\,x)\,{d \bar{G}\over dx}
$$
$$
+\left[\Big(b^{2}-{1\over 4}\Big)\,x+{a^{2}-\nu^{2}+e^{2}\over
x}+2ab+b+{1\over 2}\, {m^{2}-E^{2}+2 E e\,\sqrt{m^{2}-E^{2}} \over
m^{2}-E^{2}}\right]\bar{G}=0\,.
$$
When
$$
a = \sqrt{\nu^{2} - e^{2}} \,, \qquad b=-{1 \over 2}\;;
$$
we get
$$
x\,{d^{2}\varphi\over dx^{2}}+(2\,a+1-x)\,{d \varphi\over
dx}-\left (a-\, { E e\, \over \sqrt{m^{2}-E^{2}} }\right)\varphi=0 \;,
$$
which is a confluent hypergeometric equation
$$
x \; _{1}F_{1} '' +( \gamma -x) _{1}F_{1}' - \alpha \;_{1}F_{1} =0 \;, \qquad \alpha=
a-\, { E e\, \over \sqrt{m^{2}-E^{2}} }\,, \qquad \gamma=2a+1\,.
$$
The solutions become polynomials if $ \alpha = - n, \;n=0,1,2,...$;
 this provides us with the energy spectrum
\begin{eqnarray}
E = { m \over \sqrt{1 + e^{2} / (n + \sqrt{\nu^{2} - e^{2}})^{2}  }} \;.
\label{1.10.13c}
\end{eqnarray}
In turn, from (\ref{1.10.12}) it follows a second order equation for $F(r)$
$$
\left ( {d \over d r}  - {\nu \over r} \; \cos A + m \;\sin
A \right ) { r \over \nu \sin A + e + 2m\cos A \; r } \left ( {d
\over d r} + {\nu \over r} \;\cos A - m \; \sin A \right ) F
$$
$$
=  {\nu \sin A - e \over r} \; F \;,
$$
that is
$$
\left [ \Big( {d \over d r}  - {\nu \over r} \; \cos A + m
\;\sin A \Big)\; \Big( {d \over d r} + {\nu \over r} \;\cos A
- m \; \sin A \Big) \right.
$$
$$
\left. + \left ( {d \over dr } \ln {r \over \nu \sin A + e +
2m\cos A \; r } \right ) \Big( {d \over d r} + {\nu \over r}
\;\cos A - m \; \sin A \Big) \right.
$$
$$
\left. + { e + \nu \sin A + 2m\cos A \; r \over r}
 \; {e - \nu \sin A \over r} \right ] F = 0 \;.
$$
After simple transformations, we obtain
$$
\left [ {d^{2} \over dr^{2}} + \left ( {1 \over r} - { 2m \cos
A \over \nu \sin A + e + 2m\cos A \; r } \right ) {d \over d r} \right.
$$
$$
\left.  - { 2m \cos A \over \nu \sin A + e + 2m\cos A \; r}
\left ( {\nu \over r} \;\cos A - m \; \sin A \right ) \right.
$$
$$
\left.+ E^{2} - m^{2} + {e^{2} - \nu^{2} \over r^{2}} +
{2 e E - m \sin A \over r} \right ] F = 0 \;.
$$
With the following special notation for the additional singular point
\begin{eqnarray}
D = - {e+ \nu \sin A \over 2m \cos A} \;,
\label{1.10.14a}
\end{eqnarray}
we get the shorter form
\begin{eqnarray}
\left [ {d^{2} \over dr^{2}} + \left ( {1 \over r} - { 1 \over r - D } \right )
 {d \over d r} + {m \sin A \over r - D} - {\nu \cos A \over D} \Big({1 \over r - D}
 -{1 \over r}\Big) \right.\nonumber\\
\left.+ E^{2} - m^{2} + {e^{2} - \nu^{2} \over r^{2}} +
{2 e E - m \sin A  \over r} \right ] F = 0 \;.
\label{1.10.14b}
\end{eqnarray}
Relative to the variable $y=r/D$, this looks simpler
\begin{eqnarray}
{d^{2}F\over dy^{2}}+\left[{1\over y}-{1\over y-1}\right]\,{dF\over dy}
+\left[(E^{2}-m^{2})\,D^{2}-{\nu^{2}-e^{2}\over y^{2}} \right.\nonumber\\
\left. +{-\nu \cos A+m D\, \sin A\over y-1}+{D(2eE-m \sin A)+\nu
\cos A\over y}\right]F=0\,.
\label{1.10.14c}
\end{eqnarray}
Let $ F=y^{a}\,e^{by}\,\bar{F}(y)$; the function $\bar{F}$ satisfies
$$
{d^{2} \bar{F} \over dy^{2}}+\left({2a+1\over y}+2b-{1\over y-1}\right)\,{d \bar{F}\over dy}
$$
$$
+ \left[b^{2}+(E^{2}-m^{2})\,D^{2}+ {a^{2}-\nu^{2}+e^{2}\over
y^{2}}-{a+b+\nu \cos A-m D\, \sin A\over y-1} \right.
$$
$$
\left. + {a+b+2ab+D(2eE-m \sin A)+\nu \cos A\over y}\right]\,\bar{F}=0\,.
$$
When $a$ and $b$ are\footnote{We shall use below underlined values.}:
\begin{eqnarray}
a=\underline{+\sqrt{\nu^{2}-e^{2}}}\,, \;\; -\sqrt{\nu^{2}-e^{2}}\;,\nonumber\\
b= +\sqrt{m^{2}-E^{2}}\,D\,, \;\; \underline{-\sqrt{m^{2}-E^{2}}\,D} \;,
\label{1.10.15a}
\end{eqnarray}
this reads
\begin{eqnarray}
{d^{2} \bar{F} \over dy^{2}}+\left (2b+{2a+1\over y}-{1\over y-1}\right )\,
 {d \bar{F}\over dy}\nonumber\\
+\left ({a+b+2ab+D(2eE-m \sin A)+\nu \cos A\over y} \right.\nonumber\\
\left. - {a+b+\nu \cos A-m D\, \sin A\over y-1}\right )\bar{F}=0\,,
\label{1.10.15b}
\end{eqnarray}
which is a confluent Heun equation for $H(\alpha,\beta,\gamma,\delta,\eta,y)$
\begin{eqnarray}
H''+\left (\alpha+{1+\beta\over y}+{1+\gamma\over y-1}\right)H' \hspace{20mm}\nonumber\\
+ \left ({1\over 2}\,{\alpha+\alpha \beta-\beta-\beta
\gamma-\gamma-2\eta\over y}+{1\over 2}\, {\alpha+\alpha
\gamma+\beta+\beta \gamma+\gamma+ 2\delta+2\eta\over y-1}\right )G=0\,,
\label{1.10.15c}
\end{eqnarray}
with the parameters
\begin{eqnarray}
\alpha=2b\,, \qquad \beta=2a\,, \qquad \gamma=-2\,,
\nonumber
\\
\delta=2\,eED\,, \qquad \eta=1+m D\,\sin A-2\,eED-\nu \cos A\,.
\label{1.10.15d}
\end{eqnarray}
By imposing one of the two conditions for polynomial solutions
\cite{Ronveaux-1995}
\begin{eqnarray}
\delta=-\left (n+{\beta+\gamma+2\over 2}\right )\,\alpha\,,\qquad n=0,1,2,\dots
\label{1.10.16a}
\end{eqnarray}
we produce the energy quantization rule
$$
a= + \sqrt{\nu^{2}-e^{2}}\,,\qquad b= - \sqrt{m^{2}-E^{2}}\,D\,,
$$
$$
eED=(n+\sqrt{\nu^{2}-e^{2}})\,\sqrt{m^{2}-E^{2}}\; D \;,
$$
whence it follows
\begin{eqnarray}
E = { m \over \sqrt{1 + e^{2} / (n + \sqrt{\nu^{2} - e^{2}})^{2}  }} \;,
\label{1.10.16b}
\end{eqnarray}
which coincides with the known exact result.\par
It should be emphasized that confluent Heun equations from the cases
 1) and 2) formally coincide; however all the parameters are in fact different:
\begin{eqnarray}
1) \qquad \alpha=2b\,, \qquad \beta=2a\,, \qquad \gamma=-2\,,\nonumber\\
\delta=2\,e ER\,, \qquad \eta=1+m R\,\sin A-2\,e ER-\nu \cos A\,,
\label{1.10.17a}
\\
2) \qquad \alpha=2b\,, \qquad \beta=2a\,, \qquad \gamma=-2\,,\nonumber\\
\delta=2\,e E D \,, \qquad \eta=1+m D\,\sin A-2\,e E D-\nu \cos A\,,
\label{1.10.17b}
\end{eqnarray}
where
\begin{eqnarray}
1) \quad R = -{2e \over E + m \cos A } \;, \qquad \sin A = {e \over
\nu} \;, \qquad \cos A = \sqrt{1 - {e^{2} \over \nu^{2}} } \;;
\label{1.10.18a}\\
2) \quad D= - {e+ \nu \sin A \over 2E }\;, \qquad \cos A = {E \over m} \;
, \qquad \sin A = \sqrt{1 - {E^{2} \over m^{2}} }\;.
\label{1.10.18b}
\end{eqnarray}

\section{Standard treatment of the Coulomb problem}

It should be emphasized that the both proposed treatments of the Coulomb problem
 for Dirac equation differ from the well known one\footnote{This was firstly given by
 G. Darvin and W. Gordon (1928)) -- see in \cite{Landau-4}.}.
Let us recall this standard approach. To this end, in the radial system (\ref{1.10.1})
\begin{eqnarray}
\left ( {d \over d r } + {\nu \over r } \right ) f + \left ( E + { e \over r } +
m \right ) g = 0,\quad
\left ({d \over d r } - {\nu \over r } \right ) g - \left ( E + {e \over r } -
m \right ) f = 0\; .
\label{1.10.20'}
\end{eqnarray}

one should introduce the new functions
\begin{eqnarray}
f = \sqrt{m+E} \;(F_{1} + F_{2} )\;, \quad
g = \sqrt{m-E} \;(F_{1} - F_{2} ) \; ;
\label{1.10.20a}
\end{eqnarray}
this infers
\begin{eqnarray}
\left ( {d \over d r } + {\nu \over r } \right ) (F_{1} + F_{2} ) +
\left ( E + { e \over r } +
m \right ) {\sqrt{m-E} \over \sqrt{m +E} } \;(F_{1} - F_{2} ) = 0, \nonumber \\
\left ({d \over d r } - {\nu \over r } \right ) (F_{1} - F_{2} ) - \left ( E + {e \over r } -
 m \right ) {\sqrt{m+E} \over \sqrt{m -E} } \;(F_{1} + F_{2} ) = 0 .\nonumber
\end{eqnarray}
or,
\begin{eqnarray}
r\left ( {d \over d r } + {\nu \over r } \right ) (F_{1} + F_{2} ) +
r\; \sqrt{m^{2}-E^{2}} (F_{1} - F_{2} ) + e {\sqrt{m-E} \over \sqrt{m +E} }
\;(F_{1} - F_{2} ) = 0, \nonumber \\
 r \left ({d \over d r } - {\nu \over r } \right ) (F_{1} - F_{2} ) +
 r\; \sqrt{m^{2}-E^{2}} (F_{1} + F_{2} ) -e {\sqrt{m+E} \over \sqrt{m -E} }
 \;(F_{1} + F_{2} ) = 0 .\nonumber
\end{eqnarray}
By summing and subtracting the equations, we obtain
$$
r {d \over dr} F_{1} +\nu F_{2} + r\sqrt{m^{2}-E^{2}} F_{1} -{eE\over
\sqrt{m^{2}-E^{2}} }F_{1}-{em\over \sqrt{m^{2}-E^{2}} }F_{2} = 0
$$
$$
r {d \over dr} F_{2} +\nu F_{1} - r\sqrt{m^{2}-E^{2}} F_{2} +
{em\over \sqrt{m^{2}-E^{2}} }F_{1}+{e E\over \sqrt{m^{2}-E^{2}} }F_{2} = 0
$$
In the variables
\begin{eqnarray}
\lambda = \sqrt{m^{2}-E^{2}}, \quad x = \lambda r, \quad
{em\over \lambda } = \mu, \quad {eE\over \lambda } = \epsilon,
\label{1.10.20b}
\end{eqnarray}
these equations read shorter
\begin{eqnarray}
\left ( x {d \over dx } + x - \epsilon \right ) F_{1}
+ \left ( \nu - \mu \right ) F_{2} = 0, \quad
\left ( x {d \over dx} - x + \epsilon \right ) F_{2} +
\left ( \nu + \mu \right ) F_{1} = 0 \;.
\label{1.10.20c}
\end{eqnarray}
The system (\ref{1.10.20c}) can be solved in hypergeometric
functions \cite{Landau-4}.\par\vspace{10mm} To  detail this point,
translate  eqs. (\ref{1.10.20c}) to a new variable
 $y=2x$:
\begin{eqnarray}
\left ( y {d \over dy }  +  {y\over 2}   - \epsilon  \right )
F_{1} + \left ( \nu  - \mu  \right ) F_{2} = 0, \quad \left ( y {d
\over dy}  - {y\over 2}   + \epsilon  \right ) F_{2}   + \left (
\nu  + \mu  \right ) F_{1}  = 0 \; . \label{1.10.21}
\end{eqnarray}

\noindent From whence it follows second order differential
equations for $F_{1}$ and $ F_{2}$:
\begin{eqnarray}
\left ( y {d^{2} \over dy^{2} } + {d \over dy } +
\epsilon+{1\over 2}-{y\over 4}+{\mu^{2}-\nu^{2}-\epsilon^{2}\over
y}  \right ) F_{1} = 0\,,
\label{1.10.22a}\\
\left ( y {d^{2} \over dy^{2} } + {d \over dy } +
\epsilon-{1\over 2}-{y\over 4}+{\mu^{2}-\nu^{2}-\epsilon^{2}\over
y}  \right ) F_{2} = 0 \, . \label{1.10.22b}
\end{eqnarray}

Let us study eq.  (\ref{1.10.22a}). With the substitution
$F_{1}=y^{A}e^{By}f_{1}$ it gives
\begin{eqnarray}
\left [ y {d^{2} \over dy^{2} } + \left(2A+1+2By\right)\,{d \over
dy } +  \epsilon+{1\over 2}+B\,\left(1+2A\right)-y\left({1\over
4}-B^{2}\right)+{A^{2}+\mu^{2}-\nu^{2}-\epsilon^{2}\over y}
\right ] f_{1} = 0\,. \label{1.10.23}
\end{eqnarray}

\noindent The choice
\begin{eqnarray}
A= + \sqrt{\epsilon^{2}-\mu^2+\nu^{2}}\,,\qquad B = -{1\over 2}
\nonumber
\end{eqnarray}

\noindent simplifies eq.  (\ref{1.10.23})to that of confluent hypergeometric type
\begin{eqnarray}
\left [ y {d^{2} \over dy^{2} } + \left(2A+1-y\right)\,{d \over dy
} +  \epsilon-A  \right ] f_{1} = 0 \label{1.10.24}
\end{eqnarray}

\noindent with parameters
\begin{eqnarray}
\alpha_{1}=A-\epsilon\,,\qquad \gamma_{1}=2A+1\,. \nonumber
\end{eqnarray}
Imposing the polynomial condition $
\alpha_{1}=-n_{1} $ we  obtain the quantization rule
for  $\epsilon$:
\begin{eqnarray}
-n_{1} =-\epsilon+\sqrt{\epsilon^{2}-\mu^2+\nu^{2}}\,,
\label{1.10.25}
\end{eqnarray}

\noindent Allowing for the above definitions
\begin{eqnarray}
\lambda = \sqrt{m^{2}-E^{2}},  \quad {em\over \lambda  } = \mu,
\quad {eE\over \lambda  } = \epsilon \; ; \nonumber
\end{eqnarray}
we get
\begin{eqnarray}
\sqrt{\epsilon^{2}-\mu^2+\nu^{2}} = \sqrt{ \nu^{2}- e^{2} }\;,
\quad A = +  \sqrt{ \nu^{2}-e^{2}  }\; , \label{1.10.29}
\end{eqnarray}

\noindent and further   from  (\ref{1.10.28}) we derive a formula for energy levels
\begin{eqnarray}
{e E \over \sqrt{m^{2} - E^{2}} } =  \sqrt{ \nu^{2} -e^{2} } +
n_{2}  \equiv N_{1} \quad \Longrightarrow \quad E = {m \over \sqrt
{ 1 + e^{2} / N_{1}^{2} } } \;, \label{1.10.30}
\end{eqnarray}

\noindent which coincides with  (\ref{1.10.16b}).

Now, let us consider second equation (\ref{1.10.22b}). With the use of the substitution
$F_{2}=y^{a}e^{by}f_{2}$:
\begin{eqnarray}
\left [ y {d^{2} \over dy^{2} } + \left(2a+1+2by\right)\,{d \over
dy } +  \epsilon-{1\over 2}+b\,\left(1+2a\right)-y\left({1\over
4}-b^{2}\right)+{a^{2}+\mu^{2}-\nu^{2}-\epsilon^{2}\over y}
\right ] f_{2} = 0\,. \label{1.10.26}
\end{eqnarray}

\noindent at
\begin{eqnarray}
a=\pm\sqrt{\epsilon^{2}-\mu^2+\nu^{2}}\,,\qquad b=-{1\over 2}
\nonumber
\end{eqnarray}

\noindent we obtain an equation of the confluent hypergeometric type
\begin{eqnarray}
\left [ y {d^{2} \over dy^{2} } + \left(2a+1-y\right)\,{d \over dy
} +  \epsilon-a-1  \right ] f_{2} = 0 \label{1.10.27}
\end{eqnarray}

\noindent with parameters
\begin{eqnarray}
\alpha_{2}=a+1-\epsilon\,,\qquad \gamma_{2}=2a+1\,. \nonumber
\end{eqnarray}
Imposing the polynomial restriction
$\alpha_{2}=-n_{2} $  we get the quantization rule for
 $\epsilon$:
\begin{eqnarray}
-n_{2}=-\epsilon+1+\sqrt{\epsilon^{2}-\mu^2+\nu^{2}}\,;
\label{1.10.28}
\end{eqnarray}

\noindent so we arrive at the formula for energy levels
\begin{eqnarray}
{e E \over \sqrt{m^{2} - E^{2}} } = 1+ \sqrt{ \nu^{2} - e^{2}} +
n_{2}  \equiv N_{2} \quad \Longrightarrow \quad E = {m \over \sqrt
{ 1 + e^{2} / N_{2}^{2} } }\; . \label{1.10.30}
\end{eqnarray}

Let us find a relative coefficient between two functions:
\begin{eqnarray}
\left ( y {d \over dy }  +  {y\over 2}   - \epsilon  \right )
F_{1} + \left ( \nu  - \mu  \right ) F_{2} = 0, \nonumber
\\
\left ( y {d \over dy}  - {y\over 2}   + \epsilon  \right ) F_{2}
+ \left ( \nu  + \mu  \right ) F_{1}  = 0 \; .
\label{1.10.31a}
\end{eqnarray}
\begin{eqnarray}
F_{1}=C_{1} y^{A}e^{-y/2} F (-n_{1}, \gamma ,y) ,\quad
 F_{2}=C_{2} y^{A}e^{-y/2} F (-n_{2}, \gamma , y) , ,
 \nonumber
 \\
   A= + \sqrt{\epsilon^{2}-\mu^2+\nu^{2}}\,, \quad \gamma = 2A =1, \quad  -n_{2} = -n_{1} +1 \; .
\label{1.10.31b}
\end{eqnarray}

We substitute the expressions for the functions $F_{1},\,F_{2}$ in the first-order equation (\ref{1.10.31a})
\begin{eqnarray}
 y {d \over dy }\,\left[C_{1}\,y^{A}e^{-y/2} F (-n_{1}, \gamma ,y)\right]  +  {y\over 2} \,C_{1}\,y^{A}e^{-y/2} F (-n_{1}, \gamma ,y)  - \nonumber\\
 -
  \epsilon \,
C_{1}\,y^{A}e^{-y/2} F (-n_{1}, \gamma ,y) + \left ( \nu  - \mu  \right )C_{2}\, y^{A}e^{-y/2} F (-n_{2}, \gamma , y) = 0,
\nonumber
\label{1.10.33}
\end{eqnarray}

\noindent or
\begin{eqnarray}
 y\,C_{1}\, {d \over dy }\,F (-n_{1}, \gamma ,y) + C_{1}\,(A-\epsilon)\,F (-n_{1}, \gamma ,y)+
 \nonumber\\
     + \left ( \nu  - \mu  \right )C_{2}\,  F (-n_{1} +1, \gamma , y) = 0.
\nonumber
\label{1.10.34}
\end{eqnarray}

\noindent We apply the rule of differentiation of the confluent hypergeometric function
\begin{eqnarray}
{d \over dy }\,F (-n_{1}, \gamma ,y)=-{n_{1}\over y}\,F (-n_{1}+1, \gamma ,y)+{n_{1}\over y}\,F (-n_{1}, \gamma ,y)\,,
\nonumber
\end{eqnarray}

\noindent as a result we obtain
\begin{eqnarray}
 -C_{1}\,n_{1}\,F (-n_{1}+1, \gamma ,y)+C_{1}\,n_{1}\,F (-n_{1}, \gamma ,y) + C_{1}\,(A-\epsilon)\,F (-n_{1}, \gamma ,y)+
 \nonumber\\
     + \left ( \nu  - \mu  \right )C_{2}\,  F (-n_{1} +1, \gamma , y) = 0.
\nonumber
\end{eqnarray}

\noindent
Taking into account
$
-n_{1}=A-\epsilon\,,
$
we obtain
\begin{eqnarray}
 {C_{1}\over C_{2}}=
 { \nu  - \mu  \over n_{1}}= -{ \nu  - \mu  \over A-\epsilon}.
\label{1.10.32}
\end{eqnarray}

Now, we  substitute the expressions for the functions $F_{1},\,F_{2}$ in the first-order equation (\ref{1.10.31b})
\begin{eqnarray}
 y {d \over dy}\left[C_{2}\,y^{A}e^{-y/2} F (-n_{2}, \gamma , y)\right]  - {y\over 2}  \,C_{2}\,y^{A}e^{-y/2} F (-n_{2}, \gamma , y) +
 \nonumber\\
 + \epsilon \,C_{2}\,y^{A}e^{-y/2} F (-n_{2}, \gamma , y)
+ \left ( \nu  + \mu  \right )C_{1}\,y^{A}e^{-y/2} F (-n_{1}, \gamma ,y)  = 0 \,,
\nonumber
\end{eqnarray}

\noindent or
\begin{eqnarray}
 y\,C_{2}\, {d \over dy} F (-n_{1} +1, \gamma , y)+ C_{2}\,(A+\epsilon)\,F (-n_{1} +1, \gamma , y)-
 \nonumber\\
 -y \,C_{2}\,F (-n_{1} +1, \gamma , y)+
  \left ( \nu  + \mu  \right )C_{1}\, F (-n_{1}, \gamma ,y)  = 0 \,.
\nonumber
\end{eqnarray}

\noindent We apply the rule of differentiation of the confluent hypergeometric function
\begin{eqnarray}
{d \over dy }\,F (-n_{1}+1, \gamma ,y)=\left({-n_{1}+1\over \gamma}-1\right)\,F (-n_{1}+1, \gamma+ 1,y)+F (-n_{1}+1, \gamma ,y)\,,
\nonumber
\end{eqnarray}

\noindent
and use the formula for contiguous confluent hypergeometric functions
\begin{eqnarray}
y\,F (-n_{1}+1, \gamma+ 1,y)=\gamma F (-n_{1}+1, \gamma,y)-\gamma F (-n_{1}, \gamma,y)\,,
\nonumber
\end{eqnarray}

\noindent as a result we obtain
\begin{eqnarray}
C_{2}\, \left(-n_{1}+1-\gamma\right) F (-n_{1}+1, \gamma,y)+ C_{2}\,(A+\epsilon)\,F (-n_{1} +1, \gamma , y)-
 \nonumber\\
-C_{2}\, \left(-n_{1}+1-\gamma\right) F (-n_{1}, \gamma,y) +
  \left ( \nu  + \mu  \right )C_{1}\, F (-n_{1}, \gamma ,y)  = 0 \,.
\nonumber
\end{eqnarray}

\noindent
Taking into account
$
-n_{1}=A-\epsilon\,,
$
we obtain
\begin{eqnarray}
{C_{1}\over C_{2}} = {-n_{1}-2A \over \nu  + \mu  }={-{A+\epsilon\over \nu+\mu}}
   \,.
\label{1.10.33}
\end{eqnarray}

It is easily checked that two expressions for relative  coefficients, (\ref{1.10.32}) and (\ref{1.10.33}),
coincide. Indeed
$$
 {C_{1}\over C_{2}}=
  -{ \nu  - \mu  \over A-\epsilon}, \quad
{C_{1}\over C_{2}} =  {-{A+\epsilon\over \nu+\mu}}\;\; \Longrightarrow
$$
$$
\nu^{2}- \mu^{2} = A^{2} - \epsilon^{2}\quad  \Longleftrightarrow \quad  e^{2} \equiv e^{2} \;.
$$

%
\section{The spin $1/2$ particle in Coulomb field; solutions constructed completely by Heun functions}
The idea to construct spectra within the Heun equation theory
 seems to be a very promising one. Let us try to consider, in
 this line of arguments, the known problem of a spin $1/2$
 particle in the presence of external Coulomb field.
To this end, let us turn again to the equations in presence of Coulomb
 potential (when $\delta =+1$)
\begin{eqnarray}
\Big({d \over dr} \;+\; {\nu \over r}\;\Big) \; f \; + \; \Big( E + {e \over r} \;+ \;
 m \Big)\; g \; = \;0 \;, \nonumber \\
 \Big({d \over dr} \; - \;{\nu \over r}\;\Big)\; g \;- \; \Big( E + {e \over r}\; - \;
 m \Big)\; f\; =\; 0  \;.
\label{2.1}
\end{eqnarray}
After eliminating the function $g$, one gets
\begin{eqnarray}
{d^{2}f\over dr^{2}}+{ e \over r(E r+e +mr)}\,{df\over
dr}+\left[{e(e^{2}-\nu^{2})\over r^{2}(E r+ e+mr)}\right.\nonumber\\
\left.+{ E\,(3 e^{2}-\nu^{2})-\nu\, (m+ E)+m\,(e^{2}-\nu^{2})\over r\,
(E r+e+mr)} \right.\nonumber\\
\left. + {e\,(E+m)\,(3E-m)\over E r+ e +mr}+{r(E-m)(E+m)^{2}\over
E r+ e +mr}\right]f=0\,.
\label{2.2}
\end{eqnarray}
After changing the variable
\begin{eqnarray}
x=-{(E+m)\,r\over e }\,, \label{2.3}
\end{eqnarray}
equation (\ref{2.2}) takes the form
\begin{eqnarray}
x\,{d^{2}f\over dx^{2}}-{1\over x-1}\,{df\over dx}+\left[{ e^{2}(E
x- m x-2 E)\over E+m}+{e ^{2}-\nu^{2}\over x}-{\nu\over x-1}\right]f=0\,. \label{2.4}
\end{eqnarray}
By separating the two factors $f (x) = x^{A} e^{Cx} F (x) \;,$ one derives for $F$
\begin{eqnarray}
{d^{2}F\over dx^{2}}+\left(2C+{2A+1\over x}-{1\over
x-1}\right){dF\over dx}+\left[C^{2}+{e^{2}(E-m)\over E+m}\right.\nonumber\\
\left.+{A^{2}+ e^{2}-\nu^{2}\over x^{2}}\!+\!{A+C+2AC-2E
e^{2}/(E+m)+\nu\over x}\!-\!{A+C+\nu\over x-1}\right]\!F\!=\!0\;.
\label{2.5}
\end{eqnarray}
When $A, C$ are taken as (the bound states being of first interest)
\begin{eqnarray}
C^{2}+{e^{2}(E-m)\over E+m}=0\qquad\Rightarrow \qquad C= +
e\sqrt{m-E\over m+E}\,,\nonumber\\
A^{2}+ e^{2}-\nu^{2}=0\qquad\Rightarrow \qquad A= + \sqrt{\nu^{2}-e^{2}}\,,
\label{2.6}
\end{eqnarray}
equation (\ref{2.5}) becomes simpler,
\begin{eqnarray}
{d^{2}F\over dx^{2}}+\left(2C+{2A+1\over x}-{1\over x-1}\right){dF\over dx}\nonumber\\
+ \left[{A+C + \nu +2AC-2E e^{2}/(E +m) \over x}-{A+C+\nu\over x-1}\right]F=0\,,
\label{2.7}
\end{eqnarray}
which is the confluent Heun equation for $F (\alpha,\beta, \gamma, \delta, \eta; x)$
\begin{eqnarray}
{d^{2} \over d z^{2}} F + \left ( a + {\beta +1\over z} +
{\gamma+1 \over z-1} \right ) {d F \over d z}\nonumber\\
\hskip-0.1cm+\!\left (\!{1\over 2}{a+a\gamma+\beta+\beta \gamma+\gamma+2\delta+2\eta\over z-1}
\!+\!{1\over 2}{a\beta+a-\beta \gamma-\beta-\gamma-2\eta\over z}\right )\! F\!=\!0
\label{2.8a}
\end{eqnarray}
with parameters determined by
\begin{eqnarray}
a=2C = + 2 e \; \sqrt{m-E\over m+E} \,, \qquad \beta=2A = +2 \;
\sqrt{\nu^{2}-e^{2}} \,, \qquad\;\;\;\nonumber\\
\gamma=- 2\,, \qquad \delta= -{2E e^{2}\over E+m}\,, \qquad\eta=1-\nu+{2E e^{2}\over E+m}\,.
\label{2.8b}
\end{eqnarray}
The known condition to obtain polynomials is
\begin{eqnarray}
\delta=-a\,\left (n+{\gamma+\beta+2\over 2}\right )\,.
\label{2.9a}
\end{eqnarray}
which can bee written as
\begin{eqnarray}
-{2E e^{2}\over E+m} =- 2 e \; \sqrt{m-E\over m+E} \; ( \; n+
\sqrt{\nu^{2}-e^{2}} \; )\,, \label{2.9b}
\end{eqnarray}
or
\begin{eqnarray}
{E e \over \sqrt{m^{2} - E^{2}} } = N \;, \qquad N = n +\sqrt{\nu^{2} - e^{2}} \; ;
\label{2.9c}
\end{eqnarray}
from where it follows
\begin{eqnarray}
E = {m \over \sqrt{1 +e^{2} / N^{2}}} \;.
\label{2.10}
\end{eqnarray}

This is the exact energy spectrum for the hydrogen atom in Dirac theory.

\section {Conclusions}

In the paper, the well-known quantum mechanical problem of a spin 1/2 particle in external Coulomb
 potential, reduced to a system of two first-order differential equations, is studied from the point
 of view of possible applications of the Heun function theory to treat this system.
It is shown that in addition to the standard way to solve the problem in terms of the
 confluent hypergeometric functions (proposed in 1928 by G. Darvin and W. Gordon), there are
 possible several other possibilities which rely on applying the confluent Heun functions.
Namely, in the paper there are elaborated two combined possibilities to construct solutions:
 the first applies when one equation of the pair of relevant functions is expressed trough
 hypergeometric functions, and another constructed in terms of confluent Heun functions.
In this respect, certain relations between the two classes of functions are established.
It is shown that both functions of the system may be expressed in terms of confluent Heun functions.
All the ways to study this problem lead us to a single energy spectrum, which indicates their correctness.

\end{document}